# Nanoscale magnetic domains in polycrystalline Mn$_3$Sn films imaged by a scanning single-spin magnetometer

Senlei Li[1], Mengqi Huang[1], Hanyi Lu[1], Nathan J. McLaughlin[1], Yuxuan Xiao[2], Jingcheng Zhou[1], Eric E. Fullerton[2], Hua Chen[3,4], Hailong Wang[2,*], and Chunhui Rita Du[1,2,*]

[1]Department of Physics, University of California, San Diego, La Jolla, California 92093, USA
[2]Center for Memory and Recording Research, University of California, San Diego, La Jolla, California 92093-0401, USA
[3]Department of Physics, Colorado State University, Fort Collins, Colorado 80523, USA
[4]School of Advanced Materials Discovery, Colorado State University, Fort Collins, Colorado 80523, USA

*Corresponding author: h3wang@ucsd.edu; c1du@physics.ucsd.edu

**Keywords:** Quantum sensing, Scanning nitrogen-vacancy magnetometry, Topological magnets, Antiferromagnetic materials.

**Abstract**: Noncollinear antiferromagnets with novel magnetic orders, vanishingly small net magnetization and exotic spin related properties hold enormous promise for developing next-generation, transformative spintronic applications. A major ongoing research focus of this community is to explore, control, and harness unconventional magnetic phases of this emergent material system to deliver state-of-the-art functionalities for modern microelectronics. Here we report direct imaging of magnetic domains of polycrystalline Mn$_3$Sn films, a prototypical noncollinear antiferromagnet, using nitrogen-vacancy-based single-spin scanning microscopy. Nanoscale evolution of local stray field patterns of Mn$_3$Sn samples are systematically investigated in response to external driving forces, revealing the characteristic "heterogeneous" magnetic switching behaviors in polycrystalline textured Mn$_3$Sn films. Our results contribute to a comprehensive understanding of inhomogeneous magnetic orders of noncollinear antiferromagnets, highlighting the potential of nitrogen-vacancy centers to study microscopic spin properties of a broad range of emergent condensed matter systems.

Harnessing magnetic domain motions for advanced information processing, transfer, and storage constitutes a key mission of modern spintronic technologies.[1–4] Over the past decade, (ferri)ferromagnetic materials with net magnetization, robust stray field patterns, and prominent magneto-transport responses naturally played a leading role in this contest, and a variety of cutting-edge spintronic devices have been developed along this direction.[1–4] More recently, antiferromagnets featuring vanishingly small net magnetization and exchange-enhanced magnetic interactions emerge as a new contender of this field.[5,6] Novel functionalities such as ultrahigh-density magnetic memory,[7] improved stability,[8,9] and unconventional magnetic switching strategies[10–13] are under intensive investigation for developing next-generation, transformative micromagnetic devices. The family of noncollinear antiferromagnets $Mn_3X$ (X = Sn, Ge, Ga, Ir, Pt, Rh) are naturally relevant in this context, owing to their emergent electronic and magnetic structures. $Mn_3X$ compounds exhibit a range of exotic material properties e.g. robust anomalous Hall and Nernst effects,[14–21] chiral anomaly,[22–24] spin-momentum locking,[25] and exceptionally large spin-orbit coupling,[26,27] which provide brand new methodologies for cutting-edge spintronic applications.

A prerequisite for the noncollinear antiferromagnets $Mn_3X$ to fulfill their envisioned technological potential centers on the ability to image and control of their local spin orders at the nanoscale, which remains a formidable challenge in the current state of the art. An apparent stumbling block results from the nearly compensated magnetization, which is difficult to access by existing magnetometry techniques. While previous work has reported visualization of magnetic domains of $Mn_3Sn$ using magneto-optic Kerr effect microscopy,[13,28] the spatial resolution is fundamentally constrained by the optical diffraction limit and is constrained to the micrometer length scale. Thus, a clear picture of the microscopic magnetic textures at the nanoscale remains elusive. Here we utilize single nitrogen-vacancy (NV) scanning microscopy[29–34] to perform quantum imaging of stray field patterns of polycrystalline $Mn_3Sn$ films, and to visualize their microscopic magnetic evolution driven by external magnetic fields and electrical currents with a spatial resolution of tens of nanometers. Our results highlight the significant potential of NV centers for exploring the local spin related phenomena in noncollinear antiferromagnets, opening new opportunities for studying the interplay between external stimuli, topological magnetism, and crystalline textures in a broad range of emergent quantum materials.

We first review the basic magnetic properties of $Mn_3Sn$ and our measurement platform. $Mn_3Sn$ is a hexagonal antiferromagnet with a noncollinear inverse-triangular spin configuration.[10,27,35–40] The kagome planes of Mn atoms feature an ABAB stacking geometry as shown in Figure 1a. Ignoring the spin canting effect, the six non-collinearly ordered Mn atoms residing on two neighboring kagome layers have a nonzero cluster magnetic octupole moment despite the vanishing dipolar and quadrupolar moments.[10,14] Symmetry-allowed canting introduces an additional small remnant magnetization in the kagome planes.[10,14] For the current study, we used magnetron sputtering techniques to prepare polycrystalline $Mn_3Sn$ films with thicknesses ranging from 30 nm to 400 nm on $Al_2O_3$ and Si substrates (see Supporting Information Section 1 for details). Figure 1b shows a field-dependent magnetization curve of a 200-nm-thick $Mn_3Sn$ film measured at 300 K. The external magnetic field is applied along the out-of-plane direction of the sample and a robust canted ferromagnetic moment is observed with a magnitude that is comparable with values reported in previous literature.[41,42] The polycrystalline nature of sputter deposited $Mn_3Sn$ films is confirmed by x-ray diffraction characterizations as shown in Figure 1c, suggesting co-existing magnetic grains with kagome planes parallel and perpendicular to the substrate surface.

Robust anomalous Hall effect of prepared $Mn_3Sn$ samples has been verified by magneto-transport measurements.[43]

Quantum sensing of polycrystalline $Mn_3Sn$ films was performed using a scanning NV microscope at room temperature as illustrated in Figure 1d. A diamond cantilever[30,44,45] containing an NV single-electron spin is attached to a quartz tuning fork for force-feedback atomic force microscopy measurements. Spatial resolution of the scanning NV microscope is primarily determined by the vertical distance between the NV spin sensor and the sample surface,[30,33,46] which is set to be ~60 nm in this study. The single NV center acts as a local probe to diagnose the magnetic stray field arising from proximal $Mn_3Sn$ samples. The component of stray field along the NV axis lifts the two-fold degeneracy of NV spin energy by the Zeeman effect, which can then be optically detected by NV electron spin resonance (ESR) measurements.[30,47,48] The magnitude of the longitudinal field projection along the NV axis can be obtained from the split NV spin energy (see Supporting Information Section 2 for details). Note that individual spatial components of the stray field follow a linear dependence in the Fourier space, which allows for retrieval of its out-of-plane component $B_z$.[43,46] By scanning the diamond cantilever over the sample surface on a mesoscopic length scale, we are able to obtain nanoscale resolved stray field $B_z$ maps of Mn$_3$Sn films. In the current work, an external static magnetic field with a magnitude of ~15 G is applied during the scanning NV measurements to distinguish stray field-induced NV ESR splitting. Such a bias field is too weak to modify the local magnetic order of the Mn$_3$Sn samples studied.

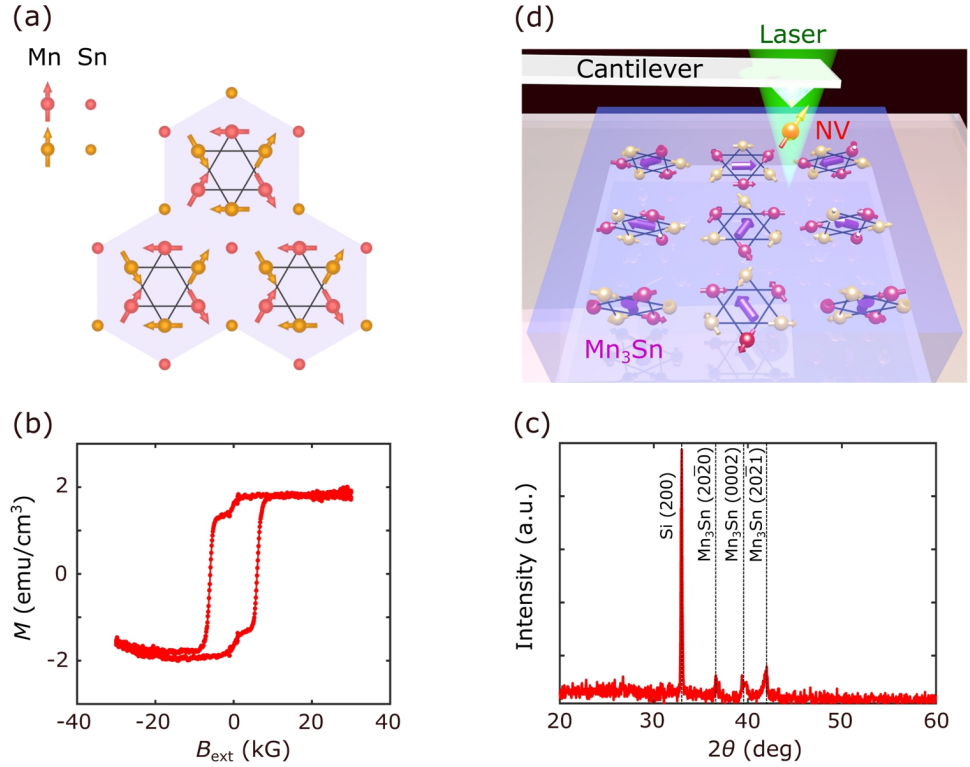

**Figure 1.** Sample characterizations and scanning NV measurement platform. (a) Schematic of the kagome lattices of $Mn_3Sn$ hosting inverse triangular spin configurations. The red and yellow arrows (balls) represent the Mn atoms and Sn atoms on two neighboring stacked kagome planes. (b) A field-dependent magnetization curve of a 200-nm-thick $Mn_3Sn$ film grown on a Si substrate measured at 300 K. The external magnetic field $B_{ext}$ is applied along the out-of-plane direction of the sample. (c) $\theta - 2\theta$ x-ray diffraction scan of the 200-nm-thick $Mn_3Sn$ film grown on a Si substrate. (d) Schematic illustration of scanning NV measurements of local spin textures of a polycrystalline $Mn_3Sn$ sample with randomly oriented kagome planes.

We now present nanoscale stray field maps of the prepared Mn$_3$Sn samples. Figures 2a-2e show our results of polycrystalline Mn$_3$Sn thin films with thicknesses of 30 nm, 50 nm, 70 nm, 100 nm, and 400 nm, respectively. All samples were pre-magnetized by a large perpendicular magnetic field [~2.5 tesla (T)] before NV measurements. Notably, multidomain signatures are observed in all the Mn$_3$Sn films. The thinner ones show fragmented magnetic domains on a length scale of tens of nanometers (Figures 2a-2c), while enhanced magnetic uniformity is observed in thicker samples possibly due to decreasing local defects and/or increase in grain sizes. The average lateral dimensions of magnetic domains formed in the Mn$_3$Sn films can be estimated by performing the autocorrelation function of presented stray field maps[43,49] (see Supporting Information Section 3 for details). Figure 2f shows that the average length scale of magnetic patterns increases with the sample thickness. The measured magnetic stray field is driven by the local canted ferromagnetic moment in Mn$_3$Sn. Microscopically, the prepared polycrystalline Mn$_3$Sn films consist of weakly coupled magnetic grains with different polarizations, resulting in stray field patterns with opposite signs between neighboring domains (see Supporting Information Section 4 for details). Due to the highly inhomogeneous magnetic orders as well as multiple magnetic easy axes in the kagome planes,[10] it is not possible to unambiguously reconstruct local magnetization patterns of the studied Mn$_3$Sn films.[46] It is instructive to note that formation of local magnetic domains in polycrystalline Mn$_3$Sn film sensitively depends on many intrinsic as well as extrinsic factors such as crystalline structure, magnetic field history, magnetic defects, local strain, and thermal cycles.[50] Thus, the observed magnetic domain patterns do not show a direct correlation to the structural/surface topography of the $Mn_3Sn$ samples.

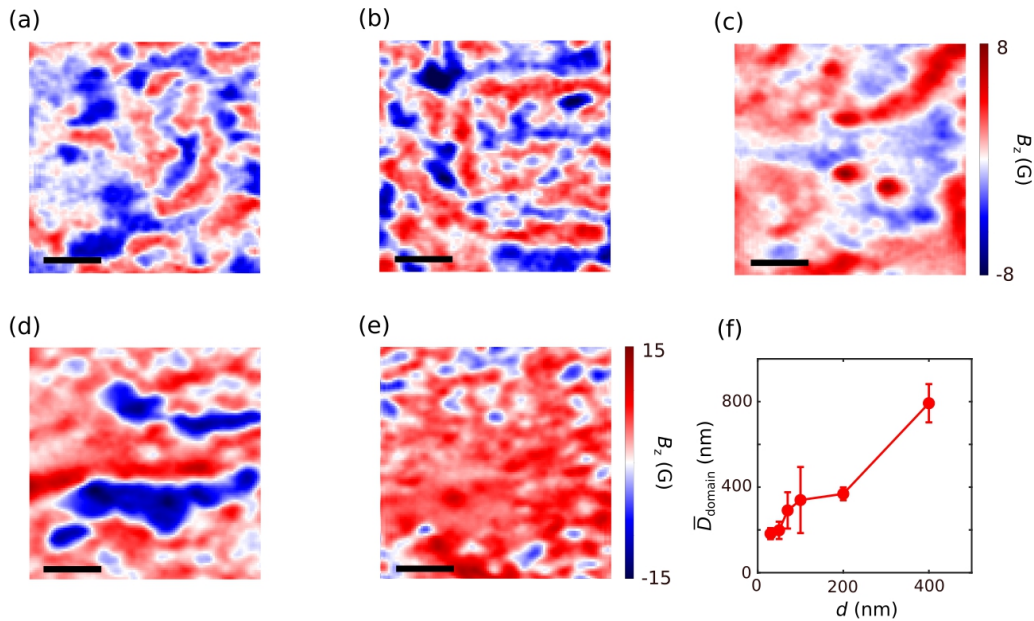

**Figure 2.** Scanning NV imaging measurement results. (a)-(e) Nanoscale stray field imaging of polycrystalline Mn$_3$Sn films with thicknesses of 30 nm (a), 50 nm (b), 70 nm (c), 100 nm (d), and 400 nm (e). Scale bar is 0.5 μm for all the images. (f) Average domain size ($\bar{D}_{\text{domain}}$) as a function of Mn$_3$Sn sample thickness $d$.

After demonstrating the operation of our scanning NV microscope, we now present data showing field-driven nanoscale magnetic reversal in a patterned $Mn_3Sn$ Hall cross device. Figure 3a shows the anomalous Hall loop of a 70-nm-thick $Mn_3Sn$ Hall device measured as a function of an out-of-plane magnetic field $B_{ext}$. The sample thickness of 70 nm is chosen by considering the balance of magnetic contrast, NV imaging quality, and electrical magnetic switching efficiency of the Mn$_3$Sn films. Anomalous Hall resistivity of the Mn$_3$Sn film is extracted to be ~2 μΩ cm, in qualitative agreement with values reported in previous studies.[10] Owing to a strong out-of-plane magnetic anisotropy, the Mn$_3$Sn sample shows almost full perpendicular remanence, offering the possibility of controlling its magnetization between two oppositely polarized magnetic states by the field training effect. Figures 3b and 3e show two stray field maps measured on the same sample area of the 70-nm-thick polycrystalline Mn$_3$Sn Hall device after field training with positive 2.5 T and negative 2.5 T perpendicular fields, respectively. After removing the external training field, the Mn$_3$Sn sample is expected to stay at the two oppositely polarized magnetic states as shown by the blue and green points in Figure 3a. One can see that the presented stray field maps (Figures 3b and 3e) largely follow the same nanoscale patterns. Zoomed-in views of two local sample areas further reveal that the polarity of individual stray field domains reverses while the domain wall boundaries remain stationary after training with oppositely polarized external magnetic field (Figures 3c, 3d, 3f, and 3g).

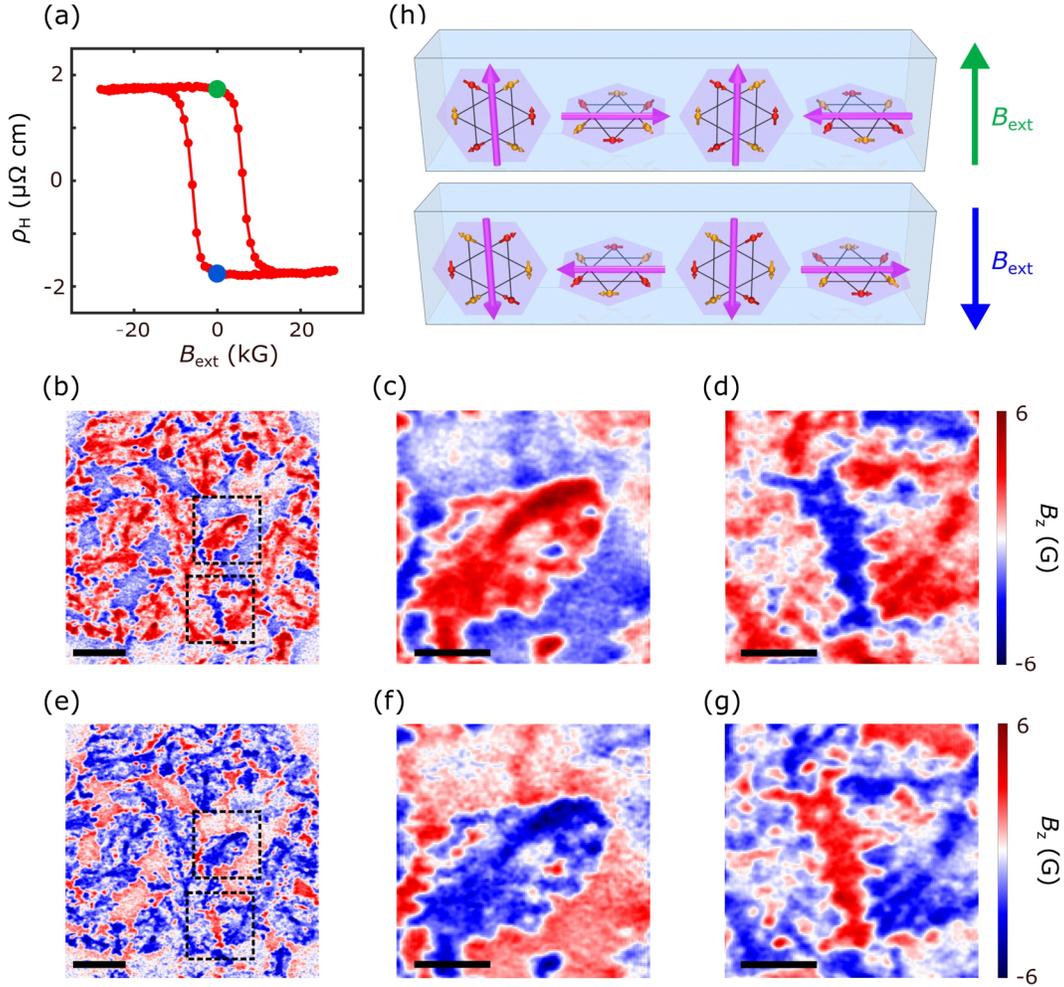

**Figure 3.** Scanning NV imaging of field-driven perpendicular magnetic switching of polycrystalline $Mn_3Sn$. (a) Anomalous Hall resistivity $\rho_H$ of a 70-nm-thick $Mn_3Sn$ sample measured as a function of a perpendicular magnetic field $B_{ext}$. The blue and green points on the hysteresis loop represent two oppositely polarized magnetic states for scanning NV measurements. (b)-(d) Stray field map (b) and zoomed-in views of two local features (c, d) of the 70-nm-thick $Mn_3Sn$ sample measured after field training with a positive 2.5 T perpendicular magnetic field. (e)-(g) The corresponding mesoscopic scale stray field map (e) and zoomed-in features (f, g) of the same sample area after field training with a negative 2.5 T perpendicular magnetic field. The squares with dashed lines in Figures 3b and 3e outline the sample area for high-resolution, fine scanning NV measurements presented in Figures 3c-3d and Figures 3f-3g. Scale bar is 2.5 μm in Figures 3b and 3e, and 0.9 μm in Figures 3c, 3d, 3f, and 3g. (h) Schematic showing magnetic order of weakly coupled $Mn_3Sn$ domains with parallel and perpendicularly oriented kagome planes under a large positive (top panel) and negative (bottom panel) magnetic training field.

The observed unconventional magnetic switching behavior results from the polycrystalline and local magnetic easy-plane nature of the Mn$_3$Sn films studied. Figure 3h shows schematics of the local magnetic order of a prepared Mn$_3$Sn film after training with a large positive (negative) perpendicular magnetic field. Polycrystalline Mn$_3$Sn samples consist of weakly coupled magnetic grains with dominantly in-plane and out-of-plane oriented kagome planes.[10,13] When reversing the sign of the external perpendicular magnetic field, magnetic grains with out-of-plane oriented kagome planes switch accordingly to follow the external field direction. Similarly, grains with their kagome planes nearly parallel to the film can also be switched by the perpendicular magnetic field, as long as the projection of the magnetic field into the kagome plane of a given grain is larger than its coercive field. Note that from the perspective of the magnetic field, the in-film-plane magnetization components of such grains are switched as well. However, these components are not mutually aligned by the magnetic field across differently oriented grains. Such spatial variations of the weak magnetization direction among magnetic grains lead to the nonuniform sign of the observed stray field maps. It also explains why training with an opposite magnetic field simply reverses the sign of the $B_z$ maps but does not change their shape. The above qualitative picture describing field-driven magnetic rearrangement of polycrystalline Mn$_3$Sn samples is also corroborated by our simulation results (see Supporting Information Section 4 for details).

In addition to the static magnetic field, current-induced spin-orbit torques (SOTs) can also be used to achieve efficient control of the local spin orders of Mn$_3$Sn.[10–13,35] Next, we utilize our scanning NV microscope to reveal evolutions of the measured stray field patterns during electrically driven perpendicular magnetic switching of Mn$_3$Sn. For these measurements, we prepared Mn$_3$Sn (70 nm)/W (7 nm) bilayer samples and patterned them to standard Hall cross devices for magneto-transport characterizations as shown in the left panel of Figure 4a. The W capping layer serves as an efficient source of spin-orbit torques (SOTs)[10,51] for driving the magnetic switching in Mn$_3$Sn, arguably together with other potential contributions such as local heating effects and inter-grain spin transfer torques.[11–13] Current driven magnetic switching measurements follow the standard procedure as reported in the previous literature.[10,43] Electrical write current pulses $I_{write}$ are applied through the current channel of a patterned Hall cross device, generating transverse spin currents through spin Hall effect[51] in the W layer, assuming the diffusive picture of spin transport is valid and spin currents are approximately well defined. The spin current then flows across the heterostructure interface and exerts SOTs on the local magnetic moments in Mn3Sn, resulting in reversible switching of the canted ferromagnetic moment with the assistance of a longitudinal bias field (see Supporting Information Section 5 for details). We have demonstrated robust electrically driven magnetic switching in the prepared Mn3Sn/W Hall devices as shown in the right panel of Figure 4a. Compared to the variation of the anomalous Hall signal during field-induced switching measurements, current-induced switching experiments achieve a

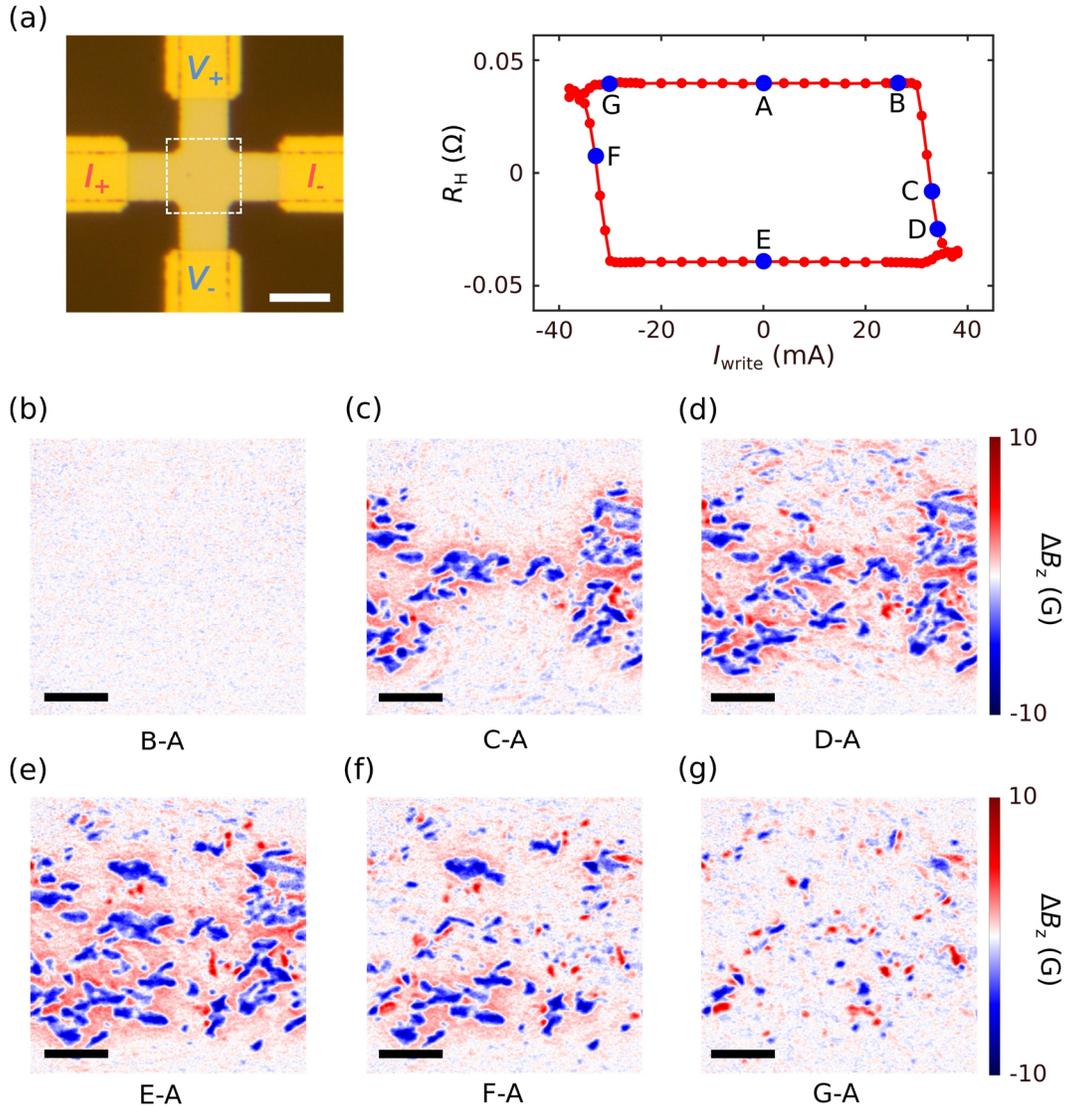

**Figure 4.** Scanning NV imaging of electrically driven magnetic switching of polycrystalline Mn$_3$Sn. (a) Left: optical image of a patterned Mn$_3$Sn/W Hall cross device with an illustration of the magneto-transport measurement geometry. The white dashed lines highlight the focused sample area for scanning NV measurements, and the scale bar is 12 μm. Right: Hall resistance $R_H$ measured as a function of write current $I_{write}$ with assistance of a longitudinal bias field. Scanning NV imaging measurements are performed at individual points from "A" to "G" marked on the current-induced magnetic hysteresis loop. (b)-(g) Scanning NV imaging of electrically driven variations of stray field ($\Delta B_z$) at different states ("B" to "G") marked on the hysteresis loop. $B_z$ map measured at the initial state magnetic "A" has been subtracted for visual clarity. Scale bar is 3 μm for all the presented stray field images.

~40% switching efficiency for polycrystalline Mn₃Sn films. At a series of points ("A" to "G") marked on the hysteresis loop, we performed scanning NV measurements after applications of individual electrical write current pulses to visualize current-driven evolution of magnetic stray field patterns at the nanoscale.

Figures 4b-4g present a series of representative stray field maps taken at the corresponding points ("B" to "G") on the current-induced magnetic hysteresis loop. For visual clarity, the stray field map measured in the initial magnetic state "A" has been subtracted to highlight the relative variations in response to the external spin stimuli. Notably, the measured field map barely changes when $I_{write}$ is below the critical value (Figure 4b). When ramping the write current above the critical value, local variations of the measured stray field start to emerge in the current leads area where the current density is higher, and then propagates to the entire Hall cross region (Figures 4c-4e). The variations of the stray field pattern become less prominent when inverting the write current

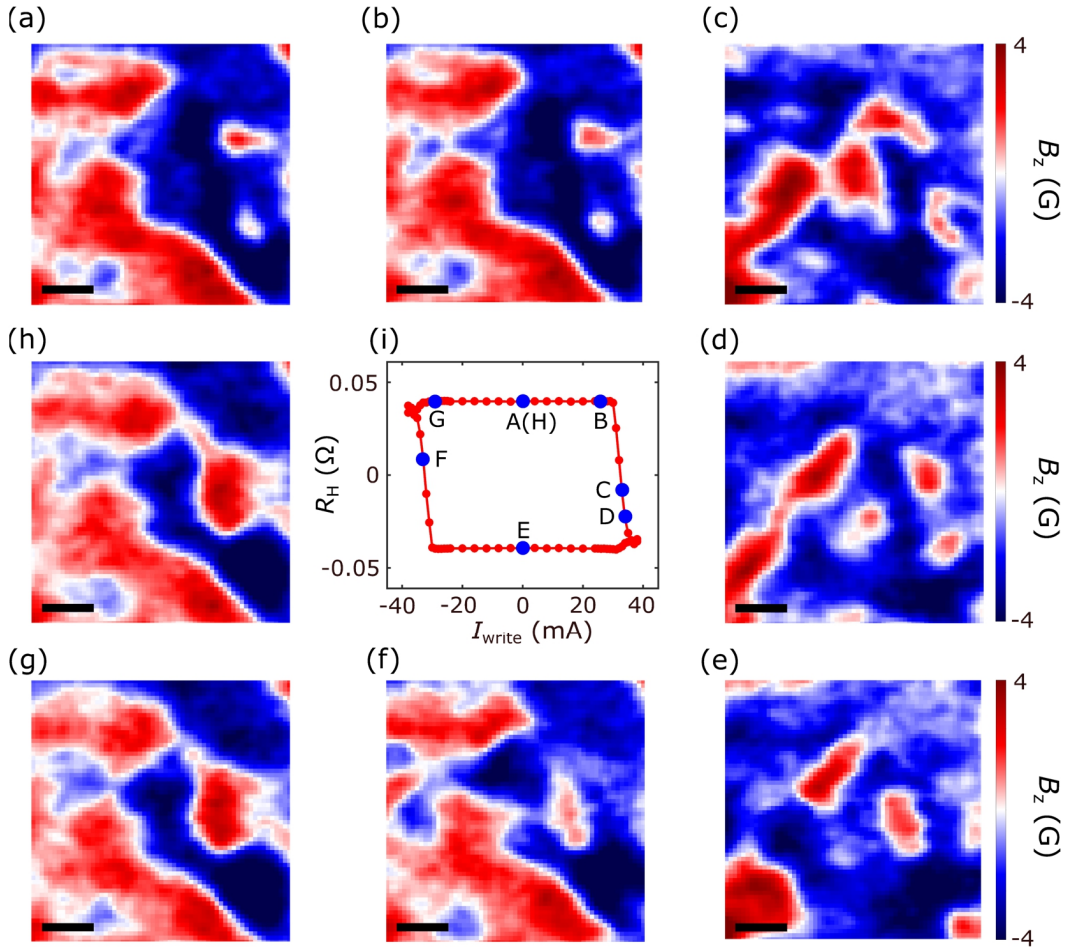

**Figure 5**. Current-driven variations of local stray field patterns at the nanoscale. (a)-(h) Nanoscale stray field imaging of electrically-induced inhomogeneous, "partial" magnetic switching in a polycrystalline Mn₃Sn/W Hall cross device. The scale bar is 0.4 μm for all the presented images. (i) Hall resistance $R_H$ of the patterned Mn₃Sn/W device measured as a function of write current $I_{write}$. A longitudinal external bias field of 500 G is applied in the electrically driven magnetic switching measurements to provide a deterministic switching polarity. Scanning NV images presented in Figures 5a-5h were performed at individual points from "A" to "H" marked on the current-induced magnetic hysteresis loop shown in Figure 5i.

into the negative regime (Figure 4f). Surprisingly, robust remnant stray field variations remain persistent after a complete electrical switching loop (Figure 4g). The presented scanning NV magnetometry results highlight a local switching behavior of weakly coupled individual $Mn_3Sn$ magnetic grains. Between the two electrically reversible magnetic states ("A" and "E"), current driven "partial" switching features rotation of the canted ferromagnetic moment in grains with perpendicularly oriented kagome planes. For grains with kagome planes parallel with the sample surface, the SOT-driven deterministic magnetic switching is not favored in this geometry.[10] To further reveal the presented microscopic features in detail, Figures 5a-5h show a zoomed-in view of current-driven evolution of stray field patterns of a local sample area measured at the corresponding magnetic states "A" to "H" on a current-induced magnetic hysteresis loop (Figure 5i). Clear inhomogeneous magnetic switching features are observed between the two oppositely polarized magnetic states "A" and "E", which is potentially attributed to the spatially varying partial switching of perpendicular $Mn_3Sn$ magnetization and/or SOT-driven rotation of in-plane oriented kagome planes (see Supporting Information Section 6 for details). Note that the presented scanning NV measurements were performed on a timescale that is much slower than the characteristic magnetic dynamics and temporal profile of applied electrical current pulses. Thus, direct visualization of spin dynamics of $Mn_3Sn$ on the nanosecond or an even faster time scale is not possible with the current measurement protocol. A clear conclusion clarifying the exact mechanism underlying electrically driven magnetic switching in $Mn_3Sn$ may not be possible based on the current study. Meanwhile, we do notice that observable difference in stray field patterns emerges after a complete current cycling loop (between the magnetic states "A" and "H"), suggesting nonreversible reconstruction of $Mn_3Sn$ magnetic structures during current-induced magnetic switching, which could be related to local thermal effects.[11,12] It is instructive to note that extended scanning NV measurements have been performed on different $Mn_3Sn$ samples to ensure the consistency of the presented results (see Supporting Information Section 6 for details).

In summary, we have demonstrated scanning NV imaging of nanoscale antiferromagnetic domains in polycrystalline $Mn_3Sn$ films. Evolution of stray field patterns arising from local $Mn_3Sn$ magnetic grains is systematically investigated in the context of external driving forces. Due to the weak inter-grain interactions, measured stray field maps of polycrystalline $Mn_3Sn$ films largely follow the same pattern for the two oppositely polarized magnetic states in field-driven magnetic switching processes. In contrast, variations of stray field maps of polycrystalline $Mn_3Sn$ samples during current-induced magnetic switching show clear inhomogeneous features together with nonreversible domain reconstruction behaviors which is potentially related to Joule heating induced local demagnetization-remagnetization processes.[11,12] Our results highlight the advantages of the "non-invasive" NV quantum metrology in both spatial and field sensitivity for studying nanomagnetism hosted by emergent condensed matter systems (see Supporting Information Section 2 for details). The current work also adds an additional ingredient to the emerging topic of $Mn_3X$ compounds, contributing to a comprehensive understanding of unconventional spin related phenomena in the family of noncollinear antiferromagnets.

**Author contributions**: S. L. performed the NV measurements and analyzed the data with M. H., H. L., and N. M. S. L. prepared and characterized the magnetic samples/devices with assistance from Y. X., E. E. F., J. Z. and H. W. H. C. provided theoretical guidance and support. C. R. D. supervised this project.

**Notes:** The authors declare no financial interest.

**Acknowledgements**: This work was primarily supported by the U.S. Department of Energy (DOE), Office of Science, Basic Energy Sciences (BES), under award No. DE-SC0022946. C. R. D., H. L. W. and M. H. acknowledge the support by the Air Force Office of Scientific Research (AFOSR) under grant No. FA9550-20-1-0319 and its Young Investigator Program under grant No. FA9550-21-1-0125. Device fabrication and characterization were partially supported by the U. S. National Science Foundation (NSF) under grant No. DMR-2046227. H. C. was supported by NSF CAREER grant No. DMR-1945023.